# One-pot Liquid-Phase Synthesis of $MoS_2$-$WS_2$ van der Waals Heterostructures for Broadband Photodetection

Shaona Bose,[1] Subhrajit Mukherjee,[1,2] Subhajit Jana,[1] Sanjeev Kumar Srivastava,[1] and Samit Kumar Ray[1,*]

[1]Department of Physics, Indian Institute of Technology, Kharagpur- 721 302, India

[2]Presently at the Faculty of Materials Science and Engineering, Technion-Israel Institute of Technology, Haifa -3203003, Israel

*Corresponding author email: physkr@phy.iitkgp.ac.in

Two dimensional (2D) van der Waals heterostructures (vdWHs) have their unique potential in facilitating the stacking of layers of different 2D materials for optoelectronic devices with superior characteristics at a reduced cost. However, the fabrication of large area all-2D heterostructures is still challenging towards realizing practical devices. In the present work, we have demonstrated a rapid yet simple, impurity free and highly efficient sonication-assisted chemical exfoliation approach to synthesize hybrid vdWHs based on 2D molybdenum disulphide ($MoS_2$) and tungsten disulphide ($WS_2$), with high yield. Microscopic and spectroscopic studies have confirmed the successful exfoliation of layered 2D materials and formation of their hybrid heterostructure. The co-existence of 2D $MoS_2$ and $WS_2$ in the vdW hybrid is established by optical absorption and Raman shift measurements along with their chemical stiochiometry determined by X-ray photoelectron spectroscopy. The spectral response of the vdWH/Si (2D/3D) heterojunction photodetector fabricated using the as-synthesized material is found to show superior broadband photoresponse compared to that shown by the individual 2D $MoS_2$ and $WS_2$ based devices. The peak responsivity is found to be ~2.15 A/W at a wavelength of ~560 nm for an applied bias of -5 V. The ease of fabrication and superior performance of the chemically synthesized vdWH-based devices have revealed their potential use for large area optoelectronic applications on Si compatible CMOS platforms.

1. Introduction

Along with graphene, which boasts of superior carrier mobility[1] and other fascinating properties[2], many 2D materials like transition metal dichalcogenides (TMDCs)[3,4], hexagonal boron nitride (hBN)[5], oxides such as $Cr_2O_3$ and $ZrO_2$[6], etc. have drawn research interests[7]. They have a layered structure where the weak interlayer van der Waals forces can be overcome by applying mild external perturbations, facilitating easy exfoliation of individual layers. The band gaps of various 2D materials range from the ultraviolet-visible (UV-vis) to the infrared (IR) region, which can be tuned by changing the number of layers in the materials. Especially, TMDCs like $MoS_2$ and $WS_2$ have gained the attention of the scientific community due to their abundance on earth, thermal stability and tunable bandgaps useful for band selective optoelectronics. As the electronic system heads towards producing aggressively scaled and low power devices, TMDCs can be an attractive choice in this regard. In bulk form, they have an indirect bandgap (for example, 1.2 eV in bulk $MoS_2$) which changes to a direct bandgap when the number of layers is reduced to a monolayer (for example, 1.9 eV in monolayer $MoS_2$)[8]. In addition, their optical absorption and emission properties are greatly influenced by the existence of excitons and exciton complexes that are of fundamental interest[9]. Atomistic $MoS_2$ and $WS_2$ devices have applications as efficient photodetectors[10], biosensors[11], thin-film transistors[12,13], energy storage[14], and memory devices[15,16] etc.

Stacking layers of different TMDCs by forming all 2D heterostructures are expected to provide wider absorption/emission bands in the system using combinatorial properties, which has been proved to be beneficial for photovoltaic[17] devices and broadband photodetection[18]. The availability of a wide range of tunable bandgaps in TMDCs allows precise stitching of different TMDCs, giving rise to a highly suitable quantum material for fabricating broadband and high speed optoelectronic devices[19–21]. As the lattice constants of $MoS_2$ and $WS_2$ are very close, they are good candidates for 2D van der Waals heterostructures (vdWHs) with the least in-plane strain by forming a type-II band alignment that results in faster charge separation across the heterostructure. While fabricating vdWHs by selectively transferring and stacking different atomistic 2D flakes onto a substrate, special attention is given to minimize the incorporation of defects and impurities at the heterointerface. Mechanically exfoliated (ME) and transferred vertical stacks of pristine TMDC layers have outstanding physical properties but offer very small active area and require complex nanolithography for the fabrication of devices[22,23]. Chemical vapour deposition (CVD) synthesizes high quality and large-area flakes with sharp and clean heterointerfaces and hence provides superior quality films; but this process involves the usage of hazardous chemical precursors and capital-intensive instrumentation[24]. In this regard, a sonication-assisted chemical exfoliation method, where the use of hazardous chemicals and sophisticated instrumentation is not a necessity, can be a way out for an easier, faster and cheaper method to synthesize vdW heterostructures. Being a versatile and low cost synthesis process, this is the main pathway to exfoliate various 2D materials like $MoS_2$[25], black phosphorus[26], etc. Furthermore, solution processed 2D heterostructures provide a high yield of flakes with variable shapes and sizes within a short period of time[27] and therefore, suitable for large area flexible device applications.

Here, we report sonochemical synthesis of $MoS_2$-$WS_2$ hybrid vdW heterostructures using ethylenediamine (EDA) – assisted chemical exfoliation in a facile manner[28]. Successful exfoliation and subsequent mixing of few layered $MoS_2$ and $WS_2$ after prolonged sonication result in multiple local heterojunctions. Atomic force microscopy (AFM), high-resolution transmission electron microscopy (HRTEM), Raman spectroscopy, and Auger electron spectroscopy (AES) have been used to evaluate the structure and quality of the heterostructures. The utilization of $MoS_2$-$WS_2$ vdWH in Si complementary metal oxide semiconductor (CMOS) compatible vertical heterojunction (in 2D/3D configuration) photodetector exhibits better current rectification and photoresponse characteristics, as compared to control devices with few-layered $MoS_2$ and $WS_2$ flakes which can be beneficial for wideband optoelectronic devices with superior performance.

## 2. Experimental section

### 2.1 Sample synthesis

Here 0.5 g $MoS_2$ and 0.5 g $WS_2$ (1:1 ratio) powder were taken separately and added to 25 ml ethylenediamine (EDA). The mixture was subjected to magnetic stirring for 24 hours at room temperature to let EDA get absorbed on the basal plane of $MoS_2$ and $WS_2$. A schematic representation of the synthesis method is shown in Figure 1. EDA acts as a chelating agent as well as a hydrogen bond donor here and it binds to the surface of the $MoS_2$ or $WS_2$ layers. This upward pull by EDA weakened the interlayer attraction and promoted layer by layer exfoliation[28]. After complete surface adsorption, the excess EDA was discarded by centrifugation and washed several times using *N,N*-dimethylformamide (DMF) to remove traces of unreacted EDA. Thereafter, the mixture was redispersed in DMF, followed by ultrasonication for two hours as shown in Figure 1(b). Reduction of interlayer attraction through hydrogen bond formation with EDA was augmented by the hydrodynamic sonochemical forces, leading to the efficient exfoliation of few layers of TMDCs. The resulting mixture containing flakes of different sizes was separated from the bulk unexfoliated part by centrifugation. These exfoliated flakes formed a colloidal suspension in DMF that was stable for a long period. Suspended $MoS_2$ and $WS_2$ layers were arranged in close proximity forming vdWHs, as presented in Figure 1(d) and 1(e). To obtain variations in the lateral dimension of the flakes, the supernatant was then centrifuged at different speeds of 2000 rpm, 5000 rpm, and 10000 rpm (named as 2k, 5k, and 10k samples, respectively), and the digital images of their colloidal dispersions are shown in Figure 1(f). The same procedure was followed for synthesizing individual 2D $MoS_2$ and $WS_2$ flakes with varying sizes to compare their properties with their hybrid vdWH.

**2.2 Characterisation**

The surface morphology of the exfoliated flakes was studied by atomic force microscope (AFM), in tapping mode [model: Agilent Technology, 5500]. The structural characterizations were carried out by high resolution transmission electron microscope (HRTEM) [model: JEOL, JEM-2100F equipped with a CCD from Gatan, Inc. USA] and X-ray diffractometer [model: Philips MRD X-ray diffractometer] using Cu-$K_\alpha$ line of wavelength 1.5418 Å. Raman spectra were recorded using Ar-Kr gas laser of 514 nm equipped with an optical microscope having 100x objective lens. Chemical composition of the samples were determined by X-ray photoelectron spectroscopy (XPS) with characteristic Al-$K_\alpha$ radiation of energy 1486.6 eV [model: PHI 5000 Versa Probe II, ULVAC PHI, INC, Japan] and Auger electron spectroscopy (AES) was done using a spectroscope [model: PHI 710, ULVAC PHI, INC, Japan]. Optical characterization was done by UV-vis-NIR spectroscopy of the as-synthesized sample, placed in a quartz cuvette with a fiber-probe based UV-vis-NIR photospectrometer, [model: Avaspec 23648]. Electrical measurements were done using Keithley semiconductor parameter analyzer [model: 4200A SCS], a broadband light source, a monochromator, and a mechanical chopper.

**2.3 Results and discussions**

A high density of exfoliated flakes has been obtained by the synthesis method, as confirmed by the TEM micrographs presented in Figure 2(a), (b) and (c), for the individual $MoS_2$ and $WS_2$ 2D nanoflakes and $MoS_2$-$WS_2$ vdWH, respectively, which were collected at 2000 rpm centrifugation speed. Prolonged ultrasonication causes the flakes to undergo size reduction in the basal plane, resulting in an average lateral size of ~100–200 nm for $MoS_2$-$WS_2$, $MoS_2$, and $WS_2$ samples, as observed in the TEM micrographs of Figure 2(a), (b) and (c) respectively. This reaffirms the efficacy of the undertaken EDA assisted sonochemical exfoliation process to exfoliate 2D TMDCs. Typical HRTEM micrographs recorded around an overlapping region of $MoS_2$ and $WS_2$ flakes present in the $MoS_2$-$WS_2$ heterostructure sample, as depicted in Figure 2(d), reveals two different sets of planes, displaced and rotated by some angle, as shown in Figure 2(e). The measured interplanar spacing of ~0.27 nm corresponds to the (100) planes of $MoS_2$[29] while that of ~0.25 nm corresponds to the (101) planes of $WS_2$[29]. The selected area electron diffraction (SAED) pattern of this overlapping region shown in Figure 2(f), exhibits two closely spaced hexagonal lattices. This confirms the retention of the hexagonal symmetry of $MoS_2$ and $WS_2$ in the vdWH, even in their exfoliated low dimensional form, preserving superior crystalline quality in the heterostructure. The exfoliation of the bulk materials have been further confirmed by AFM topography of the as-synthesized $MoS_2$-$WS_2$ vdWH obtained at 2k rpm as shown in Figure S1 of the electronic supporting information (ESI). The average thickness is found to be ~8nm, indicating the formation of few-layered flakes in the hybrid heterostructure.

The crystalline nature of $MoS_2$ and $WS_2$ in the hybrid heterostructure has been further investigated by XRD measurements at 2° grazing angle of incidence, as shown in Figure 3(a), for various samples where all the characteristic diffraction peaks are duly indexed. The intense peak at $2\theta \sim 14.6°$ for $MoS_2$, $WS_2$, and $MoS_2$-$WS_2$ vdWH is associated with the (002) planes of $MoS_2$ and $WS_2$ flakes[30,31], and it is characteristic of their hexagonal 2H phase. The dominance of this peak is due to the high density of vertically stacked 2D sheets. The full-width-half-maximum (FWHM) of this peak in the vdWH sample is 1.02° and that of individual $MoS_2$ and $WS_2$ are 1.00° and 0.9° respectively. The broadening of the FWHM for the vdWH is due to the presence of the two constituent compounds. XRD of all the samples thus clearly support the retention of crystallinity after EDA assisted solvent exfoliation.

Molecular vibrations of these samples were probed by Raman shift with a 488 nm laser. The Raman spectra of $WS_2$ presented in Figure 3(b), shows two peaks at ~359 cm$^{-1}$ and ~428 cm$^{-1}$, which are attributed to the in-plane $E^1_{2g}$ and out-of-plane $A_{1g}$ vibrational modes, respectively. The corresponding vibrational modes for $MoS_2$ nanoflakes appear at ~383 cm$^{-1}$ and ~408 cm$^{-1}$, respectively, in agreement with existing the report[32]. However, the $MoS_2$-$WS_2$ vdWH exhibits Raman peaks at ~351 cm$^{-1}$ and ~419 cm$^{-1}$ owing to the contribution from $WS_2$ and those at ~380 cm$^{-1}$ and ~407 cm$^{-1}$ are attributed to $MoS_2$ flakes. It is also to be noted that all the peaks are down-shifted in the case of the vdWH, compared to the Raman shift of individual $MoS_2$ and $WS_2$ nanoflakes, indicating the possible development of in-plane strain and hence the formation of *in situ* local heterojunction. No evolution of any new vibrational mode or unusual change in the Raman spectra of the vdWH as compared to that of bare $MoS_2$ and $WS_2$ reaffirms the crystalline quality of the exfoliated sample. In addition, the difference between the out-of-plane $A_{1g}$ and in-plane $E^1_{2g}$ modes of vibrations is found to be ~25 cm$^{-1}$ for $MoS_2$ and the ratio of the $A_{1g}$ and $E^1_{2g}$ modes for $WS_2$ is calculated to be 1.13 , both of which confirm that the 2D sheets contain only few-layers[33,] .

The investigation of the chemical state and composition of the as-synthesized vdWH has been performed using high-resolution X-ray photoelectron spectroscopy (XPS). The spectral features of Mo 3d, W 4f, and S 2p doublets have been deconvoluted and the results are shown in Figure 4. The spin-orbit split 3d orbitals of Mo, namely Mo $3d_{5/2}$ and Mo $3d_{3/2}$, are observed at ~229.5 eV and ~232.6 eV, respectively, as shown in Figure 4(a). On the other hand, as seen in Figure 4(b), two distinct peaks of W $4f_{7/2}$ and W $4f_{5/2}$ are observed at ~32.7 eV and ~35 eV, respectively. the absence of any higher energy peak for Mo and W reveal that the sample has not been oxidized to form sub-oxides of $MoO_x$ or $WO_x$ in the process of exfoliation. The binding energy values for Mo and W agree well with existing literature values for the $Mo^{+4}$ and $W^{+4}$ states[34,35]. From Figure 4(c), the S $2p_{3/2}$ and $2p_{1/2}$ peaks found at ~162.3 eV and 163.5 eV, respectively, also agree well with reports for the $S^{-2}$ state[36]. These results indicate the formation of chemical bonds like S–Mo–S and S–W–S and the absence of elemental states of M or W. The mechanism employed in synthesizing the hybrid vdWH thus has not triggered any side reaction to render these materials in any other unwanted chemical or oxidized state. Thus the chemically pure colloidal vdWH can be used to fabricate multifunctional devices. The atomic percentage of various elements present in the $MoS_2$-$WS_2$ vdWH sample has been calculated to be Mo ~ 21%, W ~ 20% and S ~ 59%. These elements are also homogenously distributed in the sample, as observed from the elemental maps produced by AES and their elemental RGB overlay presented in Figure 5. Figure 5(a) shows the field emission scanning electron micrograph of the sample on Si substrate and Figure 5(b)–(d) display the elemental maps of Mo, W, and S respectively. The overlapping region of Mo, W, and S represents the formation of the $MoS_2$-$WS_2$ vdWH, as shown in Figure 5(e). The elemental percentages of Mo, W, and S are extracted by elemental scanning over the selected area (in Figure 5(a)) presented in Figure 5(f), and they are found to be ~20.1 %, ~22.6 %, and ~57.3 %, respectively, which are in close agreement with the XPS measurements. AES elemental maps of the individual 2D $MoS_2$ and $WS_2$, along with their atomic concentrations are presented in Figures S2 and S3 (ESI). A larger fraction of S in the $MoS_2$-$WS_2$ vdWH is accounted due to the contribution from both the constituent compounds, which is in close agreement with our calculated fraction of individual elements using XPS.

Optical properties of the samples have been analyzed by the UV-vis absorption spectroscopy of $MoS_2$, $WS_2$, and the hybrid heterostructure collected after gradual centrifugation at different speeds of 2k, 5k, and 10k rpm are plotted in Figure 6. The absorption spectra for $MoS_2$ nanoflakes presented in Figure 6(a) show two distinct peaks at ~620 and ~680 nm arising from the absorptions associated with B and A excitons, respectively[37]. These peaks are caused by the formation of excitons due to electronic transitions from the spin-split valence band to the conduction band at the K-point of the Brillouin zone[37]. With an increase in centrifugation speed (i.e. rpm), the absorption peaks undergo a blue shift as shown in Figure 6(a), in accordance with the quantum confinement effect arising with gradual decrease in the dimensions of $MoS_2$ nanoflakes. Similarly, $WS_2$ nanoflakes exhibit two characteristic peaks at ~537 nm and ~644 nm ascribed to the B and A excitons, respectively, as shown in Figure 6(b). On the other hand, the $MoS_2$-$WS_2$ heterostructure shows three distinct absorption peaks at ~532 nm, ~630 nm, and ~677 nm in Figure 6(c). The feature at ~630 nm can be deconvoluted into two peaks at ~613 nm and ~637 nm, as shown in Figure S4 (ESI), due to the combined absorption of the B-exciton of $MoS_2$ and A-exciton of $WS_2$, respectively[38]. The resultant hybrid heterostructure combining 2D $MoS_2$ and $WS_2$ has a broader absorption than either of the individual materials. The co-existence of $MoS_2$ and $WS_2$ in the hybrid heterostructure thus exhibits wideband and enhanced absorption as compared to that of the constituent materials, making it a potential candidate for optoelectronic devices.

Vertical p-n junction devices have been fabricated using the as-synthesized samples to analyze their performance as a photodetector. To ensure large area vdW interactions, the hybrid heterostructure collected at a centrifugation speed of 2000 rpm, with an average lateral size of ~100–200 nm has been chosen. The $MoS_2$ and $WS_2$ nanoflakes and $MoS_2$-$WS_2$ vdWH were spin-coated onto freshly cleaned silicon substrates at 1500 rpm for a duration of 30 seconds to ensure total substrate coverage. Gold (Au) and aluminium (Al) metal electrodes were deposited on top of the film and bottom of the substrate, respectively, by thermal evaporation technique at a base pressure of ~$2\times10^{-6}$ mbar. A typical planar view of the fabricated device is shown in Figure 7(a). The effective active area of each device is ~2 $mm^2$. Control $MoS_2$ and $WS_2$ devices have also been fabricated alongside to systematically compare their performances.

Typical current–voltage (I–V) characteristics are shown in Figure 7(b)–(d) for devices made with $MoS_2$, $WS_2$ and the vdWH under dark and illumination conditions. The variation of I–V in the n-$MoS_2$/p-Si heterojunction device shows a good rectification at an applied bias of -5 V as presented in Figure 7(b), with a rectification ratio of ~$1.8\times10^2$ and moderately low dark current of ~$7.6\times10^{-6}$ A. Whereas, for the p-WS$_2$/n-Si heterojunction device, a rectification ratio of ~$3.6\times10^4$ and low dark current of ~$1.35\times10^{-8}$ A are noted as seen from Figure 7(c). However, the device with MoS$_2$-WS$_2$ vdWH/p-Si shows the maximum photo-to-dark current ratio of ~$10^3$, with a rectification ratio of ~$10^3$ and moderately low dark current of ~$2.3\times10^{-6}$ A, at -5 V applied bias, as shown in Figure 7(d). Overall, the device made of vdWH shows superior photoresponse in comparison to those made of MoS$_2$ and WS$_2$ nanoflakes individually. Therefore, the combination of MoS$_2$ and WS$_2$ provides an enhanced light harvesting capability.

The optical switching behavior of the fabricated photodetector having vdWH/Si heterojunction photodetector upon periodic illumination from a source of broadband light is shown in Figure 7(e), for different values of applied bias and it is found that the device exhibits fast photoresponse owing to the small lifespan of the photogenerated carriers. With increasing reverse bias, there is an enhanced collection efficiency of the photogenerated electrons and holes across possible MoS$_2$/WS$_2$, MoS$_2$/Si, and WS$_2$/Si heterojunctions, resulting in an increased output current. It should be noted that the possible local (2D-2D) heterojunctions, which are distributed throughout the device, have a cumulative effect on the output current, leaving no control over a single (2D-2D) heterojunction. A maximum of ~120 times enhancement in the value of current upon illumination is noted at -3 V bias, in the vdWH/Si device. Response time of the photodetector has been calculated by measuring the time taken by the output current to rise from 10% to 90% of its maximum value. The average response time ($\tau_{Res}$) of our vdWH/Si photodetector, estimated from the magnified view of one optical switching cycle at -3 V, as shown in Figure 7(f), is found to be ~50 ms. Hence, the photoresponse of the device fabricated using the vdWH is moderately sharp and can be employed for various optoelectronic devices.

Spectral responsivity of a photodetector is a measure of the change in output current with incident radiation. The responsivity profiles for various photodetectors have been recorded with a broadband light source and the results are presented in Figure 8(a). The vdWH/Si photodetector yields a much higher responsivity over a range of 400–800 nm, compared to the individual $MoS_2$/Si or $WS_2$/Si heterojunction devices. The vdWH/Si device covers the spectral range of both $MoS_2$ and $WS_2$ based devices, and thus it has an extended spectral range of operation in the visible region with superior responsivity. The NIR contribution to the responsivity around ~1020 nm is due to the intrinsic absorptions in the underlying Si substrate and this response remains almost unchanged for all the devices. A maximum responsivity of ~2.15 A/W (at -5 V) is noted over a range of 550–670 nm in the vdWH device. This operational range of wavelengths is closer to the bandgap of $MoS_2$ and $WS_2$ (~1.8–2.2 eV), leading to higher absorption of photons. In the NIR region, it has a peak value of ~0.97 A/W for an applied bias of -5 V, at ~1020 nm. In comparison, the peak responsivities for $MoS_2$/Si and $WS_2$/Si devices are found to be ~1.6 A/W and ~1.04 A/W respectively, in the visible region, which are much lower than that of the vdWH/Si device. The specific detectivity of a device reflects how efficiently the device can detect the required signal in the presence of a background noise. The $MoS_2$-$WS_2$ vdWH/Si device is found to exhibit the highest detectivity of ~$2\times10^{11}$ Jones at ~560 nm, as compared to the devices fabricated using the individual layers, as shown in Figure 8(b). For comparison, commercial silicon based photodetector, Newport, Model 818-BB-21, has a peak responsivity of 0.47 A/W[39]. The optical characteristics is thus found to be enhanced in the vdWH device, as concluded in table 1.

**Table 1**: Comparison of various device parameters for $MoS_2$, $WS_2$ and $MoS_2$-$WS_2$ vdWH devices

| Device materials | Dark current (A) | Rectification | Photo-to-dark current ratio | Peak Responsivity (A/W) | Peak Detectivity (Jones) |
|---|---|---|---|---|---|

| | | | | | | |
|---|---|---|---|---|---|---|
| MoS$_2$ | 7.6 × 10$^{-6}$ | 1.8 × 10$^2$ | 10$^2$ | 1.7 (660 nm) | 1.63 × 10$^{11}$ (660 nm) |
| WS$_2$ | 1.35 × 10$^{-8}$ | 3.6 × 10$^4$ | 10$^2$ | 1.0 (600 nm) | 9.6 × 10$^{11}$ (600 nm) |
| MoS$_2$-WS$_2$ vdWH | 2.3 × 10$^{-6}$ | 10$^3$ | 10$^3$ | 2.1 (560 nm) | 2.05 × 10$^{11}$ (560 nm) |

The mechanism of generation and collection of charge carriers in the MoS$_2$-WS$_2$ vdWH/Si heterojunction can be explained with the help of an energy band diagram presented in Figure 8. The energy levels (conduction and valence band edges) of various materials[25] before they are brought together are shown in Figure 8(c). After they are joined together, the energy levels align themselves in such a way that a type-II heterojunction between MoS$_2$ and WS$_2$ is formed as shown in Figure 8(d) and this is favorable for movement of charge carriers in the device. Under a reverse bias, the depletion region at a p-n junction widens, facilitating generation of more photocarriers. As the applied reverse bias is increased, it helps faster separation of charges across the junction, with a corresponding re-alignment of energy levels, as shown in Figure 8(e). When light is incident across the depletion region with energy higher than the band gap of the materials, electron-hole (e-h) pairs are generated in all the possible MoS$_2$/WS$_2$, MoS$_2$/Si and WS$_2$/Si heterojunctions. These electrons and holes move in the opposite direction under an applied external bias and get collected in the external circuit as photocurrent. The presence of both MoS$_2$ and WS$_2$ increases the range of wavelengths of photons that can be absorbed by the vdWH device. Greater the width of the extended depletion regions encountered by photons, higher is the rate of e-h pair generation, hence higher is the photocurrent. Therefore, this study reveals the enormous potential of chemically synthesized MoS$_2$-WS$_2$ vdWHs for applications in photosensing devices.

## 3. Conclusions

We have demonstrated a simple and easy and inexpensive, EDA-assisted sonochemical exfoliation of 2D $MoS_2$ and $WS_2$ layers to form hybrid vdWH without altering their crystal structure and phase. Various structural, spectroscopic, and microscopic characterizations have confirmed the presence of both $MoS_2$ and $WS_2$ 2D sheets in the vdWH and the formation of local heterojunction. The heterojunction (vdWH/Si) photodetector shows moderately good rectification behavior and sharp photoresponse under broadband illumination. The spectral response from the photodetector fabricated using $MoS_2$-$WS_2$ vdWH has shown superior responsivity over an extended spectral range of 400–800 nm with combined response from $MoS_2$ and $WS_2$ nanoflakes. A maximum responsivity of ~2.15 A/W in the visible range of 550–670 nm and detectivity of ~$2\times10^{11}$ Jones at ~560 nm have been achieved from the device. This study shows the potential use of chemical synthesis routes for synthesis of hybrid vdW heterostructure and their application in broadband photodetection.

## 4. Acknowledgements

The authors duly thank Dr. Arup Ghorai, IIT Kharagpur for his help with sample preparation and useful discussions. The authors are also thankful to the Central Research Facility, IIT Kharagpur and XPS facility (DST-FIST) in the Department of Physics, IIT Kharagpur for the access to laboratories.

## 5. Conflict of Interest

The authors declare no conflict of interest.

## 7. Figures

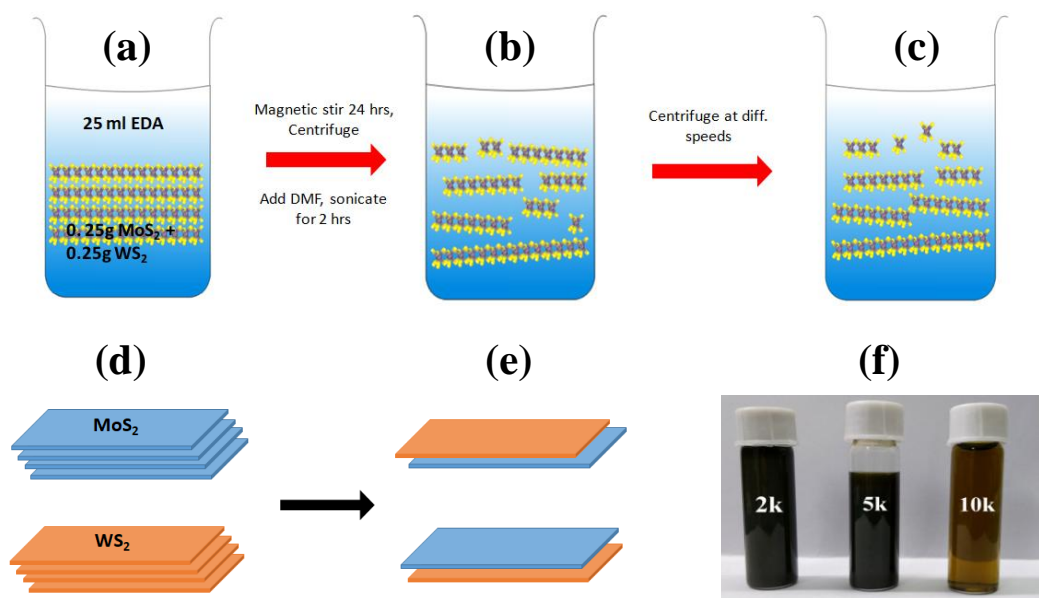

**Figure 1**: Schematic representation of the different steps of solvent exfoliation: (a) Bulk powders of $MoS_2$ and $WS_2$ were added to EDA solution, (b) After magnetic stirring for 24 hours, EDA was washed out and the mixture of $MoS_2$ and $WS_2$ was re-dispersed into fresh DMF and sonicated for 2 hours and (c) centrifugation at different rotational speeds to extract few layers of exfoliated $MoS_2$-$WS_2$ vdWH. Schematic representation of this vdWH formation is shown in (d). After detachment from bulk body, (e) $MoS_2$ is placed on top of $WS_2$ layers or vice versa. (f) Digital photograph of as-synthesized $MoS_2$-$WS_2$ vdWH.

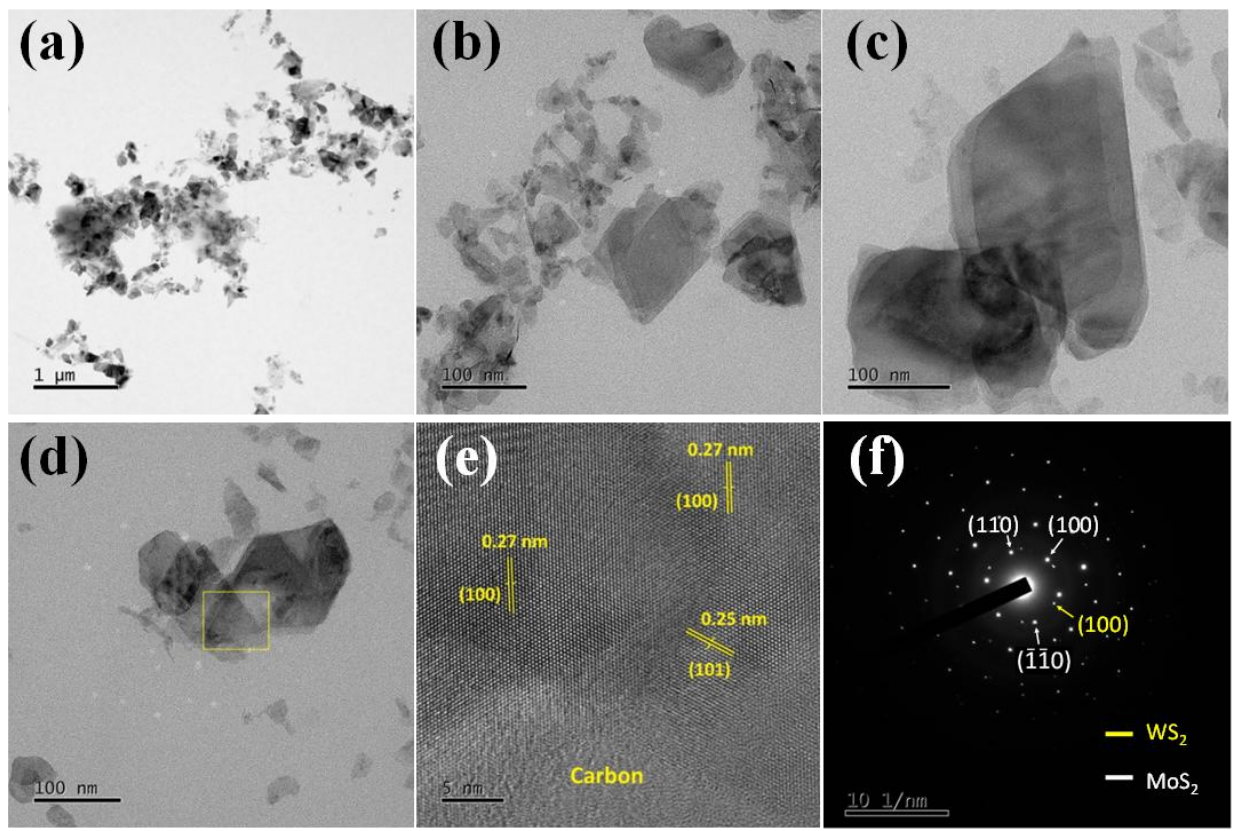

**Figure 2**: HRTEM micrographs of (a) $MoS_2$-$WS_2$ vdWH over a large area, (b) $MoS_2$, and (c) $WS_2$ flakes. (d) Heterostructure sample showing overlapping region of two flakes. (e) High-resolution TEM showing lattice fringes from two different lattices in the overlapping region and (f) SAED pattern representing the crystallinity of the vdWH.

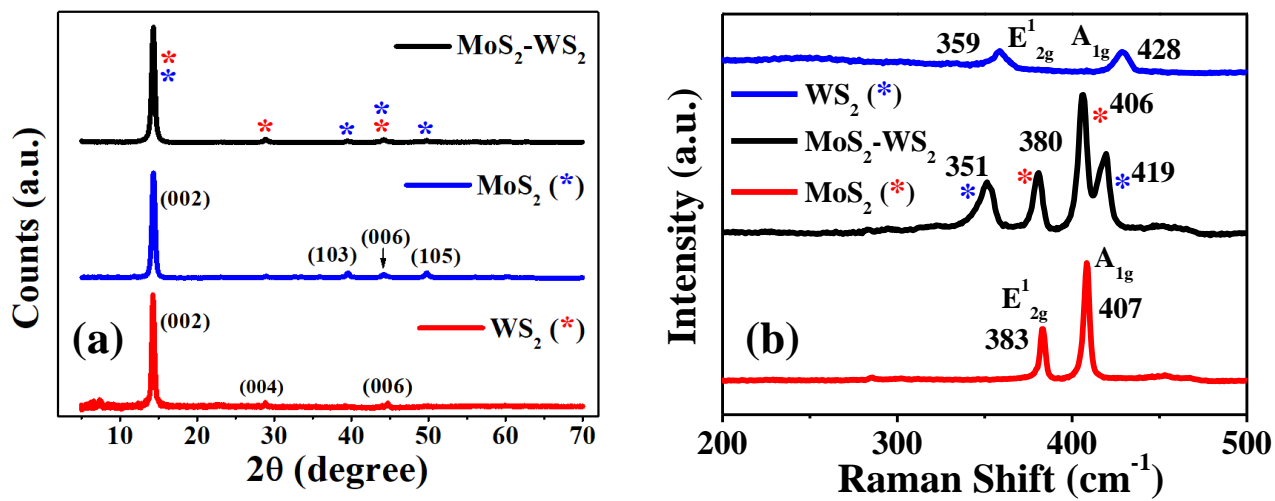

**Figure 3**: (a) XRD pattern of $MoS_2$, $WS_2$ and their heterostructure collected at 2k rpm centrifugation speed. (b) Raman spectra of $MoS_2$, $WS_2$ and $MoS_2$-$WS_2$ vdWH (2k rpm).

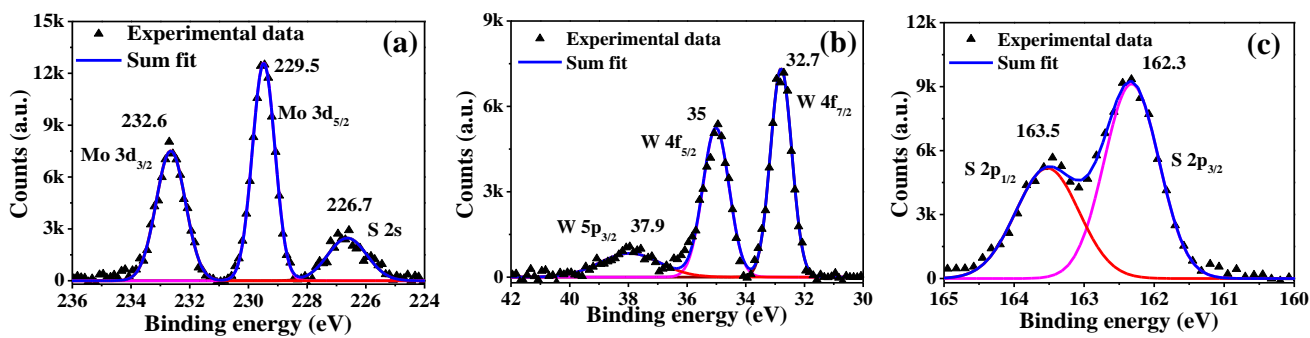

**Figure 4**: XPS core level spectra of (a) Mo, (b) W, and (c) S of the as-synthesized $MoS_2$-$WS_2$ vdWH.

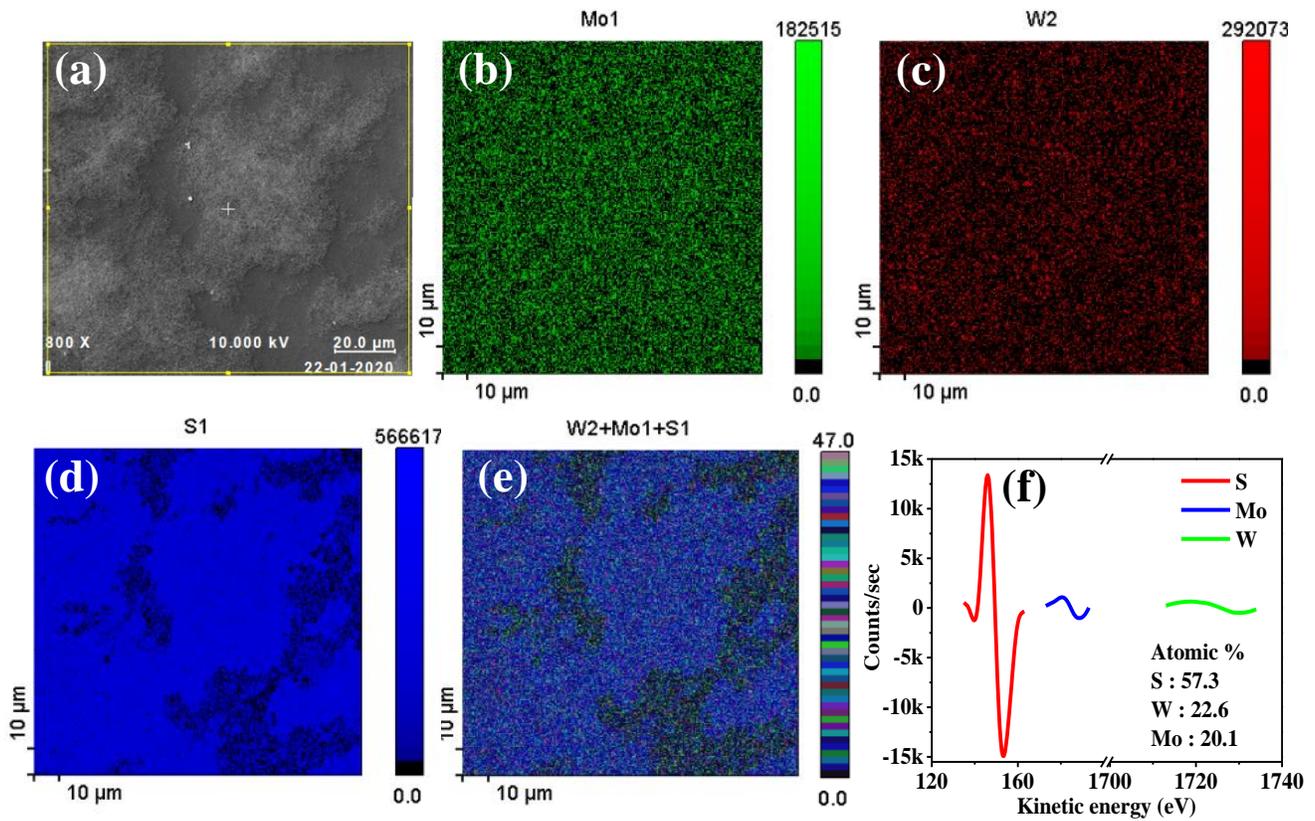

**Figure 5**: (a) SEM micrograph of the vdW heterostructure used for Auger electron mapping. AES maps for elemental distribution of (b) Mo, (c) W, (d) S and (e) their combination. (f) Elemental percentages of the above mapped elements are: 57.3% of S, 22.6% of W, and 20.1% of Mo.

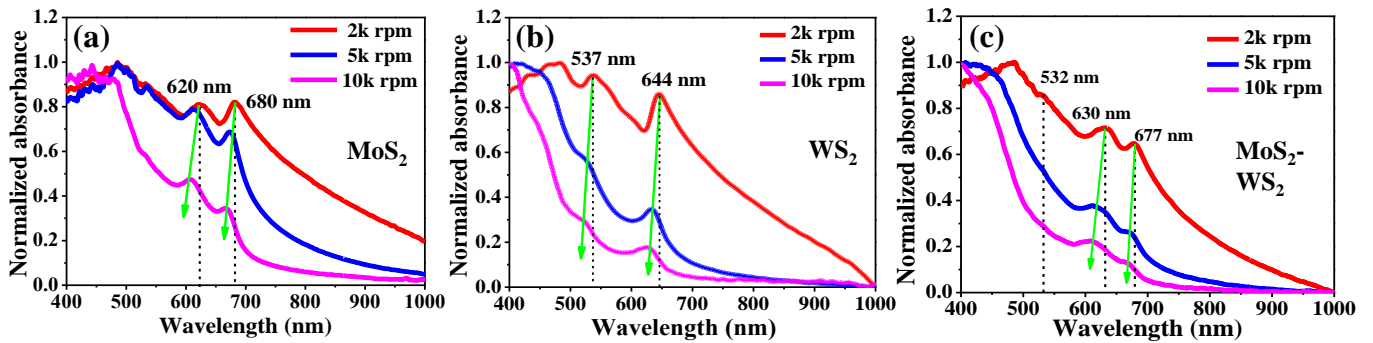

**Figure 6**: UV-vis absorption spectra of the as-synthesized (a) $MoS_2$, (b) $WS_2$ and (c) $MoS_2$-$WS_2$ vdWH.

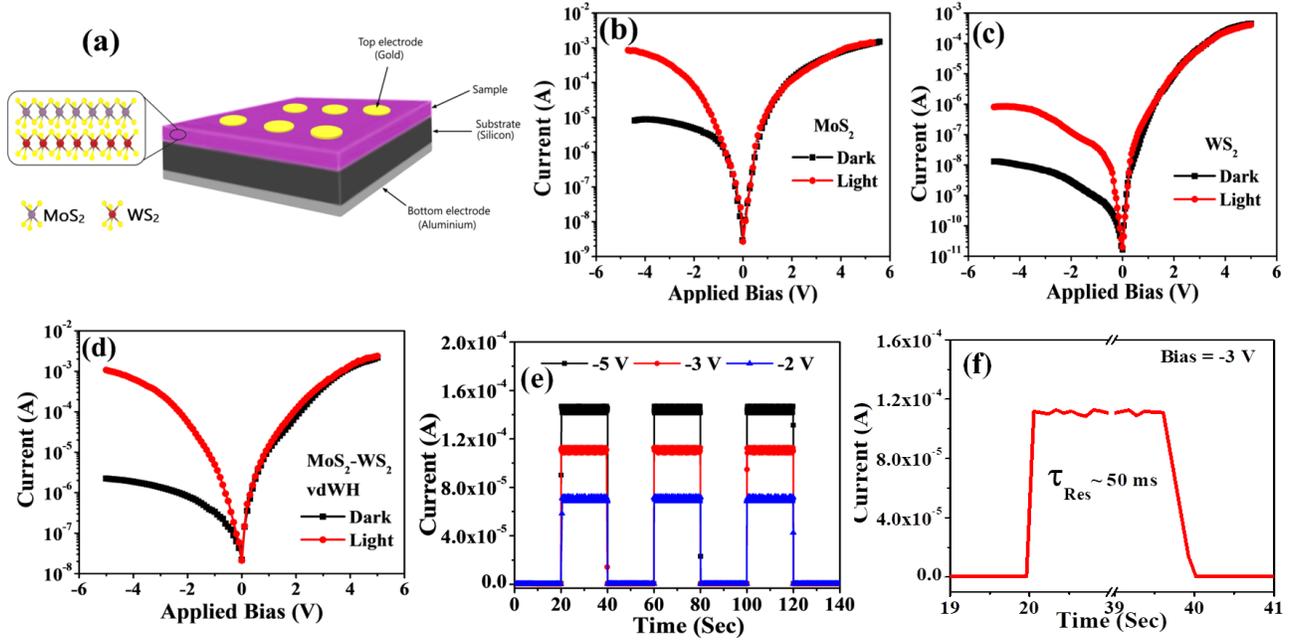

**Figure 7**: (a) Schematic representation of the fabricated heterojunction device. Typical I-V characteristics under dark and illuminated conditions in (b) n-MoS$_2$/p-Si, (c) p-WS$_2$/n-Si and (d) MoS$_2$-WS$_2$ vdWH/p-Si devices. (e) Optical modulation characteristics of MoS$_2$-WS$_2$ vdWH/Si heterojunction device at different biases for incident broadband light and (f) zoomed-in view of a switching cycle.

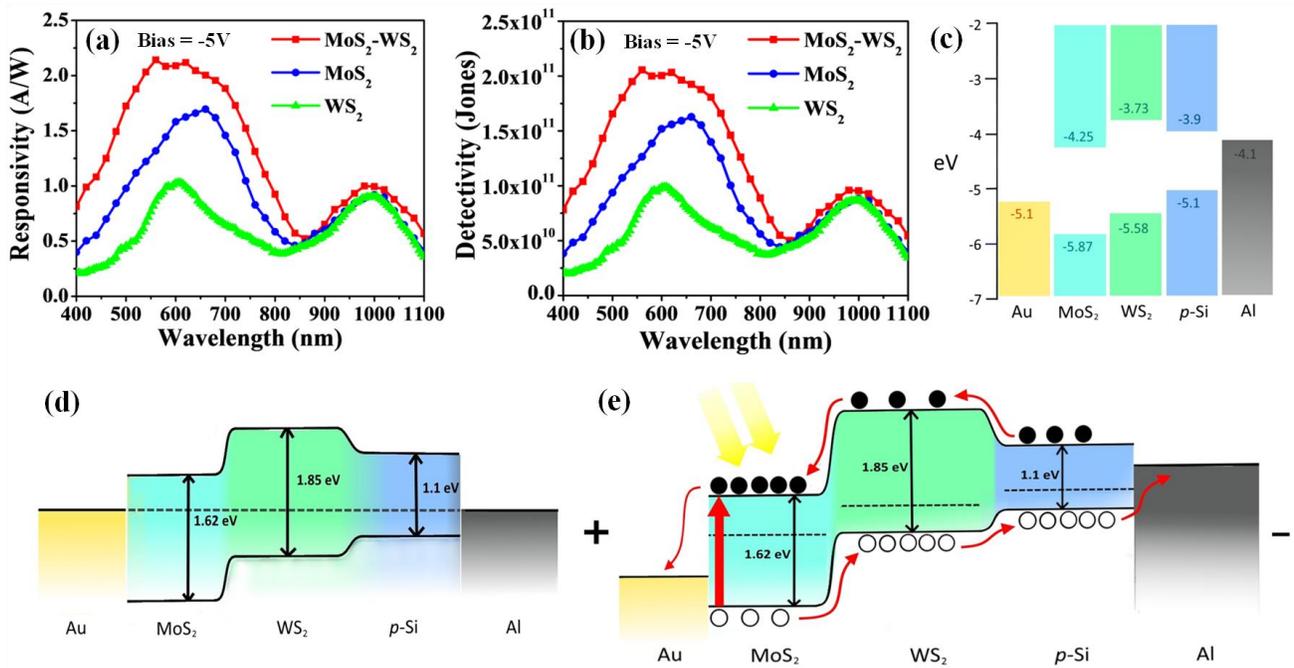

**Figure 8**: (a) Responsivity and (b) Detectivity of the $MoS_2$, $WS_2$ and vdW heterostructure devices over a wide range of wavelengths in the visible to near-IR region. (c) Band edges of various materials used in the device. Energy band diagram of heterostructure and mechanism of photodetection of $MoS_2$-$WS_2$ vdWH/Si under (d) dark and equilibrium condition and (e) illuminated and biased condition.

# Electronic supporting information

# One-pot Liquid-Phase Synthesis of $MoS_2$-$WS_2$ van der Waals Heterostructures for Broadband Photodetection

Shaona Bose,[1] Subhrajit Mukherjee,[1,2] Subhajit Jana,[1] Sanjeev Kumar Srivastava,[1] and Samit Kumar Ray[1,*]

[1]Department of Physics, Indian Institute of Technology, Kharagpur- 721 302, India

[2]Presently at the Faculty of Materials Science and Engineering, Technion-Israel Institute of Technology, Haifa -3203003, Israel

*Corresponding author email: physkr@phy.iitkgp.ac.in

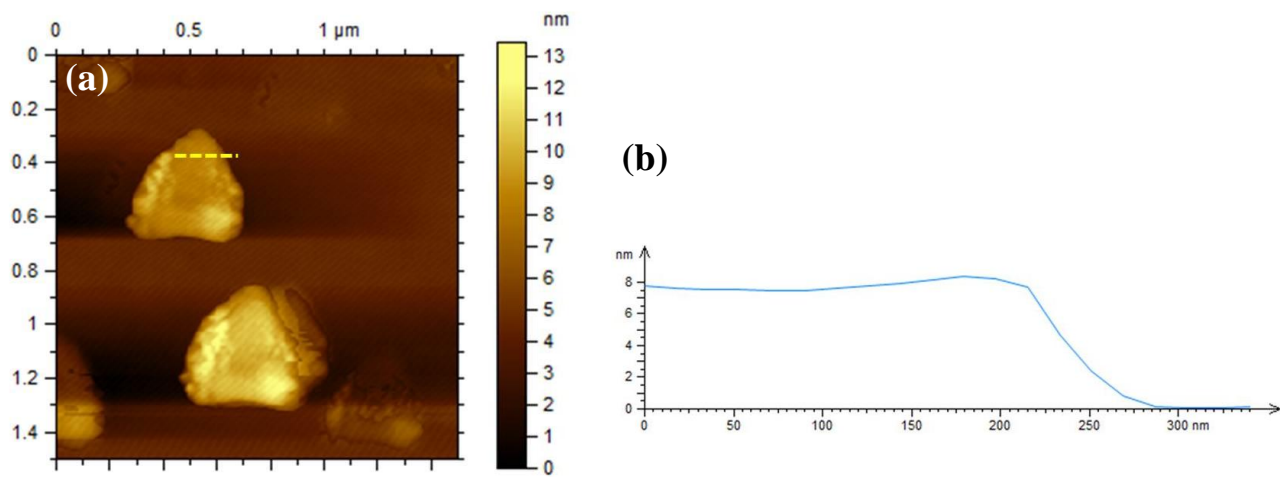

**Figure S1**: (a) Surface topography of $MoS_2$-$WS_2$ vdWH as recorded using AFM and (b) respective height profile along dashed line, showing that the average thickness is ~8 nm for vdWHs (samples collected at 2000 rpm centrifugation speed).

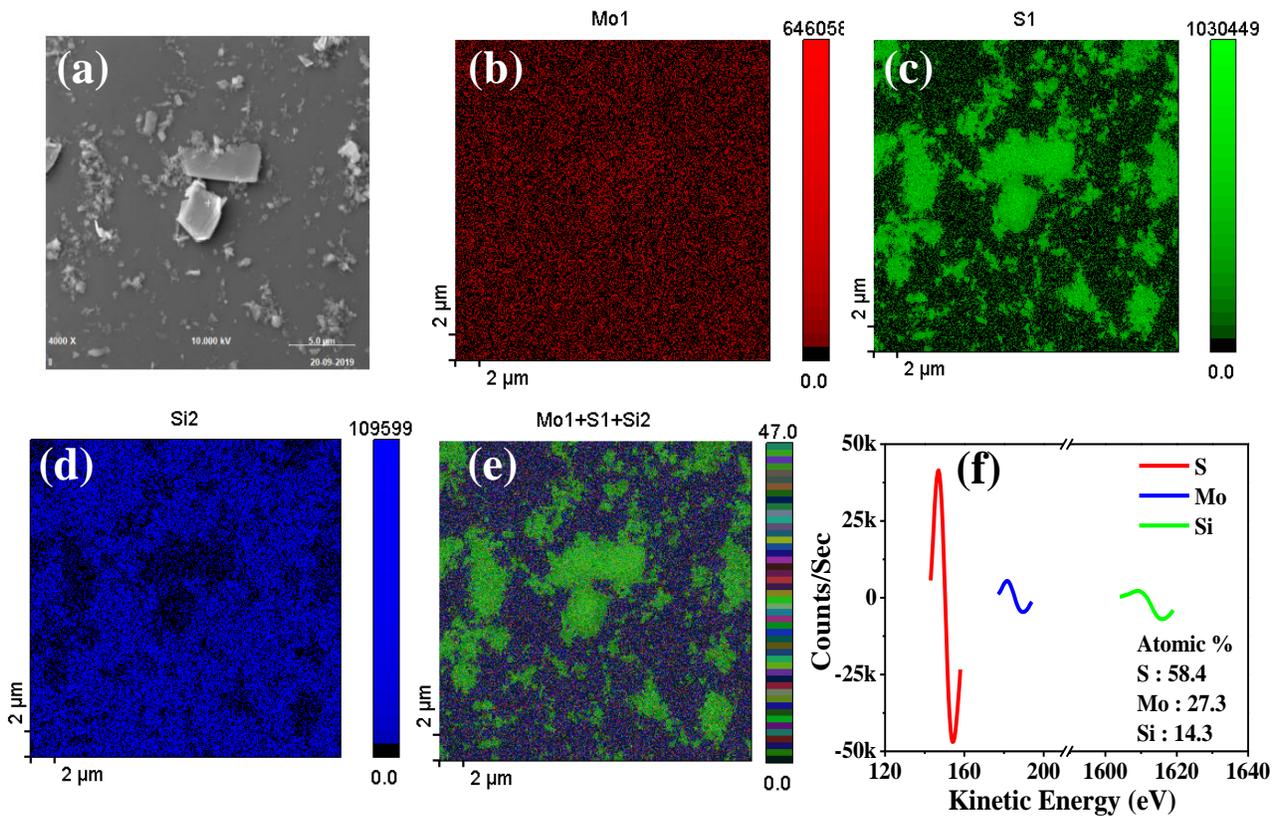

**Figure S2**: (a) SEM micrograph of MoS$_2$ sample used for Auger electron mapping. AES map for elemental distribution of (b) Mo, (c) S, (d) Si and (e) their combination. (f) Percentage elemental content of the sample.

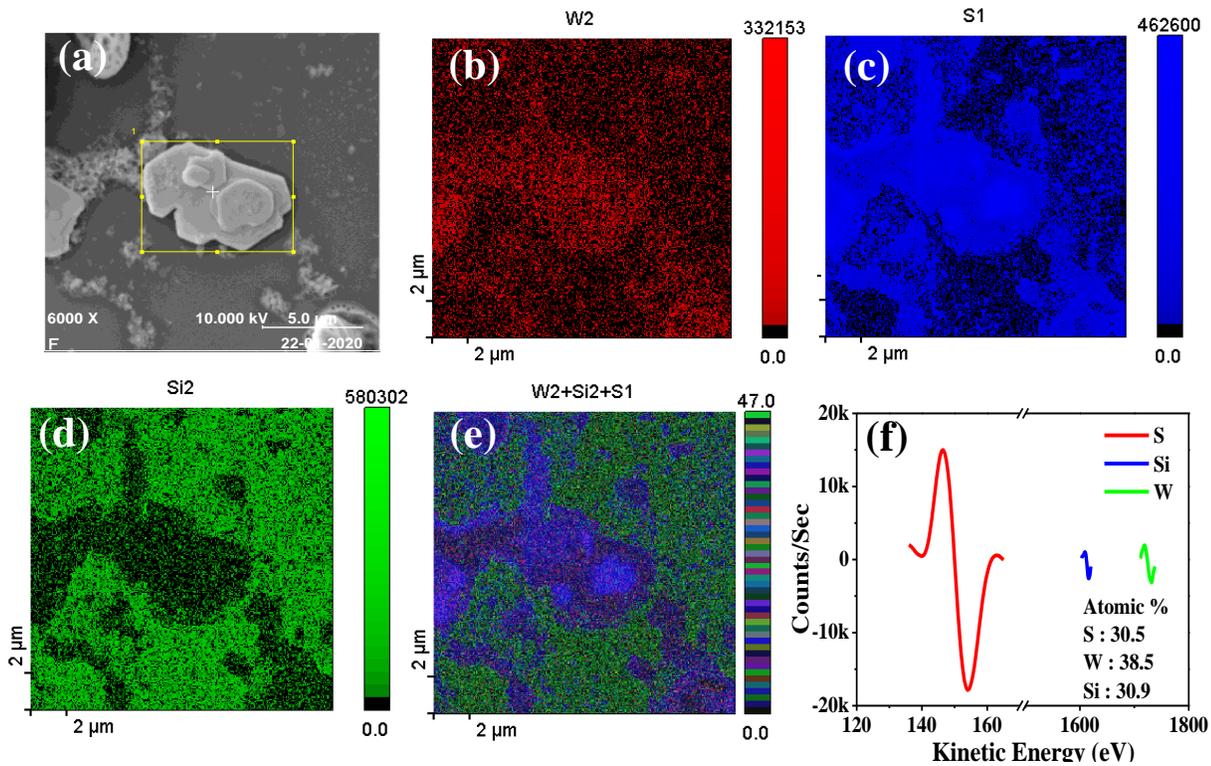

**Figure S3**: (a) SEM image of WS$_2$ sample used for Auger electron mapping. AES map for elemental distribution of (b) Mo, (c) S, (d) Si and (e) their combination. (f) Percentage elemental content of the sample.

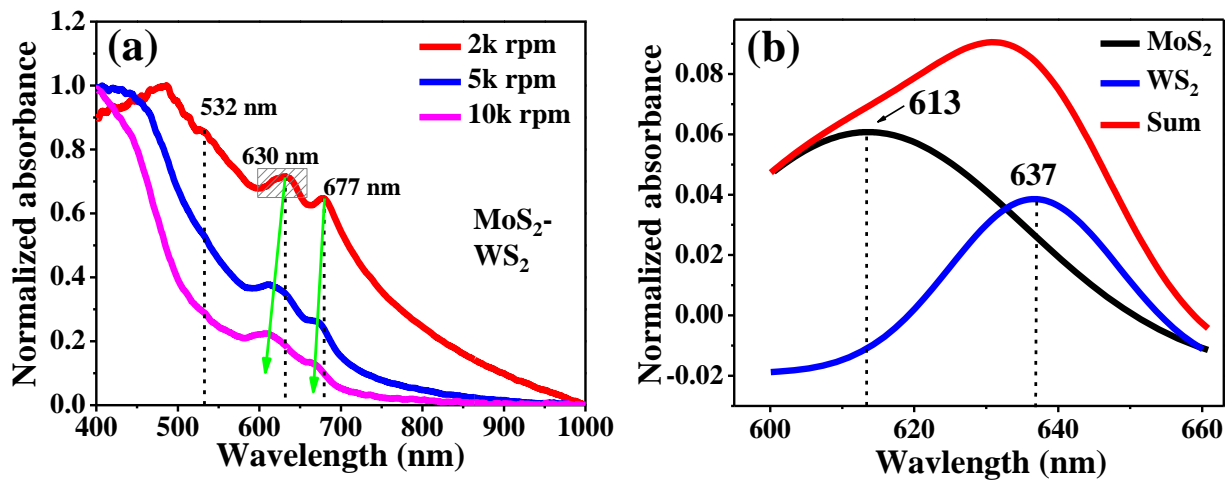

**Figure S4**: (a) UV-vis absorption spectra of the as-synthesized MoS$_2$-WS$_2$ vdWH. (b) Deconvolution of the shaded portion in (a).